\documentclass[12pt]{iopart}
\usepackage{amssymb,accents}

\pdfoutput=1

\newcommand{\wt}{\tilde}
\newcommand{\wh}{\hat}
\newcommand{\wb}{\bar}

\renewcommand{\th}[1]{\wh{\wt{#1}}}
\newcommand{\hb}[1]{\wb{\wh{#1}}}
\newcommand{\bt}[1]{\wt{\wb{#1}}}
\newcommand{\thb}[1]{\wh{\wt{\wb{#1}}}}
\newcommand{\wg}{\grave}

\newcommand{\cQ}{\mathcal{Q}}
\newcommand{\cF}{\mathcal{F}}
\newcommand{\oT}{\mathsf{T}}

\begin{document}
\paper{A multidimensionally consistent version of Hirota's discrete KdV equation}
\author{James Atkinson}
\address{School of Mathematics and Statistics, The University of Sydney, NSW 2006, Australia.}
\date{27.12.11}
\begin{abstract}
A multidimensionally consistent generalisation of Hirota's discrete KdV equation is proposed, it is a quad equation defined by a polynomial that is quadratic in each variable.
Soliton solutions and interpretation of the model as superposition principle are given.
It is discussed how an important property of the defining polynomial, a factorisation of discriminants, appears also in the few other known discrete integrable multi-quadratic models.
\end{abstract}

\section{Introduction}
This paper mainly concerns the following polynomial that is quadratic in each of four variables:
\begin{equation}
\eqalign{
\fl\quad \cQ_{p,q}(v,\wt{v},\wh{v},\th{v}):=[p(v\wh{v}+\wt{v}\th{v})-q(v\wt{v}+\wh{v}\th{v})]^2\\-4[ (v-\th{v})(\wt{v}-\wh{v})+p-q ][ (v-\th{v})(\wt{v}-\wh{v}) + (p-q)v\wt{v}\wh{v}\th{v} ].}\label{model}
\end{equation}
In the simplest setting it defines an equation $\cQ_{p,q}(v,\wt{v},\wh{v},\th{v})=0$ that relates variables on vertices of each quad on the regular $\mathbb{Z}^2$ lattice, whence $v=v(n,m)$, $\wt{v}=v(n+1,m)$, $\wh{v}=v(n,m+1)$ and $\th{v}=v(n+1,m+1)$ are values of the dependent variable as a function of independent variables $n,m\in\mathbb{Z}$, and $p,q\in\mathbb{C}$ are constant parameters of the equation. 
In the case $p=-q=2$ (\ref{model}) is reducible,
\begin{equation}
\cQ_{2,-2}(v,\wt{v},\wh{v},\th{v})= 16(v\wt{v}\wh{v}-\wt{v}\wh{v}\th{v}-\wt{v}+\wh{v})(v\wh{v}\th{v}-v\wt{v}\th{v}-\th{v}+v).\label{reduced}
\end{equation}
Each polynomial factor in (\ref{reduced}) defines a quad equation recognisable as Hirota's discrete KdV \cite{hir3}, and it is natural to view (\ref{model}) as a generalisation of that equation with the feature of being multidimensionally consistent.

The multidimensional consistency identified in \cite{nw,bs1} is an elegant algebraic integrability feature for fully discrete equations.
It is a property inherent in systems that emerge as the superposition principle for commuting auto-B\"acklund transformations, and therefore models exhibiting the property can be found throughout soliton and geometric transformation theory.
Most are in the form of rational systems that define single-valued evolution from appropriate initial data.
A notable exception is a model obtained originally by Kashaev in \cite{kas} which relates variables on vertices of a cube in $\mathbb{Z}^3$ by a {\it multi-quadratic} polynomial, we call it the discrete CKP equation because of its interpretation, due to Schief \cite{sch}, as superposition principle for the CKP hierarchy.
Probably the only documented example of a quad equation with similar features is the discrete KdV equation obtained by Adler and Veselov \cite{av},
\begin{equation}
\eqalign{
\fl\quad
[(u-\th{u}+p+q)(\wh{u}-\wt{u}+p-q)-p^2+q^2]^2\\-4(\wt{u}-\wh{u})[(\wt{u}+\th{u})(u-\th{u})p-(\wh{u}+\th{u})(u-\th{u})q+(\wt{u}-\wh{u})pq]=0,}
\label{avmodel}
\end{equation}
which is the superposition principle for B\"acklund transformations of the KdV equation.
This model was also derived by Kassotakis and Nieszporski in \cite{kn} through an approach which allows explicit verification of its multidimensional consistency.
As described in \cite{av} it is related to the equation
\begin{equation}
(w-\th{w})(\wh{w}-\wt{w})=p-q,\label{lpkdv}
\end{equation}
which is superposition principle for B\"acklund transformations of the {\it potential} KdV equation \cite{we,nc}, by the system
\begin{equation}
u+\wt{u}+p=(w-\wt{w})^2, \quad u+\wh{u}+q=(w-\wh{w})^2. \label{kdvbt}
\end{equation}
In the continuous setting $u=\partial_x w$ and (\ref{kdvbt}) is the auto-B\"acklund transformation for the potential KdV equation, however in the fully discrete setting the system (\ref{kdvbt}) is viewed as a B\"acklund transformation connecting the distinct equations (\ref{avmodel}) and (\ref{lpkdv}).
The multi-affine model (\ref{lpkdv}) is much better studied than its multi-quadratic counterpart (\ref{avmodel}), this is natural because it is simpler and the models are anyway connected due to the B\"acklund transformation (\ref{kdvbt}).
Of course it is possible that the class of integrable multi-quadratic models, of which (\ref{avmodel}) is a member, contains systems that are not B\"acklund related to multi-affine ones.
For instance such transformation connecting model (\ref{model}) to a multi-affine equation is not presently known.

The relationship between the equation defined by (\ref{model}) and the model (\ref{avmodel}) is actually hierarchical, with (\ref{avmodel}) appearing by the following degeneration procedure
\begin{equation}
\fl\quad
\lim_{\epsilon\rightarrow 0}\frac{1}{4\epsilon^4}\cQ_{2i\epsilon p,2i\epsilon q}(\sqrt{i}(1+\epsilon u),\sqrt{i}(1+\epsilon \wt{u}),\sqrt{i}(1+\epsilon \wh{u}),\sqrt{i}(1+\epsilon \th{u}))=0,
\end{equation}
where $i^2=-1$, as a limiting sub-case of the equation defined by (\ref{model}).

The contents of this paper are as follows.
The multidimensionality of the new example of multi-quadratic model (\ref{model}) is discussed first in terms of a consistent system of polynomials.
An important additional property, namely {\it discriminant factorisation}, is observed in Section \ref{DF}, it removes the apparent awkwardness of dealing with multivalued evolution of the quadratic models and allows straightforward verification of the multidimensional consistency.
This turns out to be connected to the concept of vertex-bond variables discussed by Hietarinta and Viallet in \cite{hv}.
Soliton solutions of the equation defined by (\ref{model}), which are strikingly similar to the well-known Hirota discrete KdV soliton solutions, are given in Section \ref{SS}.
In Section \ref{HM} property (\ref{reduced}) is used to further clarify the relationship between these models.
In the final section we discuss the importance of the discriminant factorisation property in the context of other known multi-quadratic discrete integrable models.

\section{Multidimensionality}\label{MD}
\begin{figure}[t]
\begin{center}
\begin{picture}(200,140)
\input{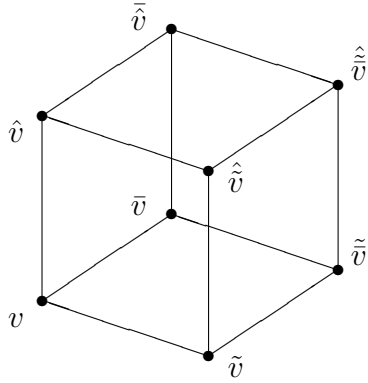}
\end{picture}
\end{center}
\caption{\label{cubepic}Variables assigned to the vertices of a cube. System (\ref{cubesys}) determines four possible values of $\thb{v}$ from initial data $v$, $\wt{v}$, $\wh{v}$ and $\wb{v}$.}
\end{figure}

For consistency in multidimensions the key properties of the polynomial (\ref{model}) are first the symmetry
\begin{equation} 
\cQ_{p,q}(v,\wt{v},\wh{v},\th{v}) = \cQ_{q,p}(v,\wh{v},\wt{v},\th{v}), \label{covar}
\end{equation}
and second the consistency of the system
\begin{equation}
\eqalign{
\cQ_{p,q}(v,\wt{v},\wh{v},\th{v})=0, \quad& \cQ_{p,q}(\wb{v},\bt{v},\hb{v},\thb{v})=0,\\
\cQ_{q,r}(v,\wh{v},\wb{v},\hb{v})=0, \quad& \cQ_{q,r}(\wt{v},\th{v},\bt{v},\thb{v})=0,\\
\cQ_{r,p}(v,\wb{v},\wt{v},\bt{v})=0, \quad& \cQ_{r,p}(\wh{v},\hb{v},\th{v},\thb{v})=0,
}\label{cubesys}
\end{equation}
for any choice of the parameters $p$, $q$ and $r$, which is usually visualised by assigning variables to vertices of a cube as in Figure \ref{cubepic}, and equations to faces.
Due to the quadratic defining polynomial the consistency needs to be understood by considering the three polynomial equations that remain after the elimination of variables $\th{v}$, $\hb{v}$ and $\bt{v}$ from (\ref{cubesys}).
Direct calculation shows that the remaining polynomials share one irreducible factor which is degree four in the remaining variables $v$, $\wt{v}$, $\wh{v}$, $\wb{v}$ and $\thb{v}$.
This means that for generic initial data $\{v,\wt{v},\wh{v},\wb{v}\}$ the system (\ref{cubesys}) determines four possible values for $\thb{v}$.
An expression for $\thb{v}$ in terms of this data will be given in the following section.

These two basic properties, namely covariance and consistency around a cube, imply consistency of the $d$ dimensional system
\begin{equation}
\cQ_{p_i,p_j}(v,\oT_iv,\oT_jv,\oT_i\oT_jv)=0, \quad i,j\in\{1\ldots d \},\label{mds}
\end{equation}
where now $v$ is defined on $\mathbb{Z}^d$, $\oT_1\ldots\oT_d$ are shift operators and $p_1\ldots p_d$ are corresponding {\it lattice parameters} associated with each direction.
This is the wider and more natural setting for the model defined by polynomial (\ref{model}).
Initial data for (\ref{mds}) can be specified on co-ordinate axes, but, unlike for the multi-affine defining polynomial, such data does not determine a unique solution.
However this multi-valued evolution can brought under control due to a further important property of the defining polynomial.

\section{Discriminant factorisation}\label{DF}
It is straightforward to verify that the discriminants of (\ref{model}) are reducible, specifically
\begin{equation}
\fl\quad
\Delta[\cQ_{p,q}(v,\wt{v},\wh{v},\th{v}),\th{v}] = 16(p-q)^2(\wt{v}-\wh{v})^2(1+pv\wt{v}+v^2\wt{v}^2)(1+qv\wh{v}+v^2\wh{v}^2),
\label{discrim}
\end{equation}
and due to its symmetry there is a similar formula for discriminants of this polynomial with respect to the other variables $v$, $\wt{v}$ and $\wh{v}$.

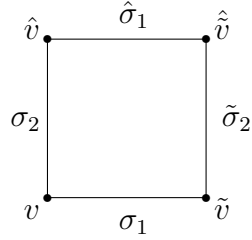
\begin{figure}[t]
\begin{center}
\begin{picture}(100,100)(0,0)
\multiput(21,12)(60,0){2}{{\line(0,1){60}}}
\multiput(21,12)(0,60){2}{{\line(1,0){60}}}
\put(7,39){{$\sigma_2$}}
\put(48,0){{$\sigma_1$}}
\put(87,39){{$\wt{\sigma}_2$}}
\put(48,78){{$\wh{\sigma}_1$}}
\put(21,12){\circle*{3}}
\put(81,12){\circle*{3}}
\put(21,72){\circle*{3}}
\put(81,72){\circle*{3}}
\put(12,4){$v$}
\put(84,4){$\wt{v}$}
\put(12,72){$\wh{v}$}
\put(84,72){$\th{v}$}
\end{picture}
\end{center}
\caption{Variables assigned to the vertices of a quadrilateral, and auxiliary variables to the edges.}
\label{quadpic}
\end{figure}

The form of (\ref{discrim}) suggests introduction of auxiliary variables $\sigma_1$ and $\sigma_2$ satisfying equations
\begin{equation}
\sigma_1^2 = 1+pv\wt{v}+v^2\wt{v}^2, \quad \sigma_2^2 = 1+qv\wh{v}+v^2\wh{v}^2,\label{dvdef}
\end{equation}
which are naturally associated with edges of the lattice in the $n$ and $m$ directions respectively, see Figure \ref{quadpic}.
By solving equation $\cQ_{p,q}(v,\wt{v},\wh{v},\th{v})=0$ for $\th{v}$ and exploiting these variables we obtain an equation of the form
\begin{equation}
\cF_{p,q}(v,\wt{v},\wh{v},\th{v},\sigma_1\sigma_2)=0
\end{equation}
which is by construction of polynomial degree one in $\th{v}$ and $\sigma_1\sigma_2$, for instance:
\begin{equation}
\fl\quad\cF_{p,q}(v,\wt{v},\wh{v},\th{v},\sigma_1\sigma_2):=v[p(v\wh{v}-\wt{v}\th{v})+q(v\wt{v}-\wh{v}\th{v})]-2(v-\th{v})(\sigma_1\sigma_2-1-v^2\wt{v}\wh{v})\label{Fdef}
\end{equation}
(the precise form of this equation is chosen because it is simple, but it is not unique).
Sequentially solving for each of the quad variables in the same way leads to the following system
\begin{equation}
\eqalign{
\cF_{p,q}(v,\wt{v},\wh{v},\th{v},\sigma_1\sigma_2)=0,\\
\cF_{p,q}(\wt{v},v,\th{v},\wh{v},\sigma_1\wt{\sigma}_2)=0,\\
\cF_{p,q}(\wh{v},\th{v},v,\wt{v},\wh{\sigma}_1\sigma_2)=0,\\
\cF_{p,q}(\th{v},\wh{v},\wt{v},v,\wh{\sigma}_1\wt{\sigma}_2)=0,
}
\label{quadsys}
\end{equation}
up to the choice of sign of the discriminant terms appearing in the last argument.
Each equation in (\ref{quadsys}) implies $\cQ_{p,q}(v,\wt{v},\wh{v},\th{v})=0$ modulo the relations (\ref{dvdef}).
The choice of sign for the discriminant terms is important for self-consistency, ensuring that any one equation in (\ref{quadsys}) is a consequence of the other three.
Within this constraint other choices of the signs can also break covariance or consistency on the cube, so we restrict attention to this canonical one.

The main benefit of replacing the quad equation defined by (\ref{model}) with system (\ref{quadsys}) is the single-valued evolution from initial data.
The usual initial value problem for a quad-equation involves specifying the dependent variable on vertices along some admissible lattice path or collection of paths.
The modification required for system (\ref{quadsys}) is that variables on path edges, subject to the constraint (\ref{dvdef}), also need to be specified.
For instance on the single quad of Figure \ref{quadpic} the variables $\{\th{v},\wh{\sigma}_1,\wt{\sigma}_2\}$ are determined rationally by (\ref{quadsys}) from $\{v,\wt{v},\wh{v},\sigma_1,\sigma_2\}$.
This system is therefore a model of vertex-bond type as described in \cite{hv}, but with the additional feature of preserving algebraic relations (\ref{dvdef}) on edges.
In this setting the multi-valued evolution of the underlying correspondence is reflected in the non-uniqueness of initial edge variables when they are defined in terms of vertex variables by (\ref{dvdef}).

As mentioned, the discriminant signs have been chosen in system (\ref{quadsys}) so that it inherits covariance from (\ref{model}), this enables it to be imposed on each face of the cube in Figure \ref{cubepic} without requiring to specify orientation.
Initial data to check consistency on the cube for (\ref{quadsys}) is $\{v,\wt{v},\wh{v},\wb{v},\sigma_1,\sigma_2,\sigma_3\}$ subject of course to the edge relations: (\ref{dvdef}) and $\sigma_3^2 = 1+rv\wb{v}+v^2\wb{v}^2$.
Putting system (\ref{quadsys}) in place of each equation in (\ref{cubesys}) leads by direct calculation to the following equation:
\begin{equation}
\fl\quad\eqalign{
& v(\wb{v}\thb{v}-\wt{v}\wh{v})r(p-q)^2-2(\wb{v}-\thb{v})(\sigma_1\sigma_2-1-v^2\wt{v}\wh{v})(r-p)(q-r)\\
+& v(\wt{v}\thb{v}-\wh{v}\wb{v})p(q-r)^2-2(\wt{v}-\thb{v})(\sigma_2\sigma_3-1-v^2\wh{v}\wb{v})(p-q)(r-p)\\
+& v(\wh{v}\thb{v}-\wb{v}\wt{v})q(r-p)^2-2(\wh{v}-\thb{v})(\sigma_3\sigma_1-1-v^2\wb{v}\wt{v})(q-r)(p-q) = 0.
}
\label{cubesol}
\end{equation}
This determines $\thb{v}$ uniquely in terms of the initial data on the cube. 
Transformations $(\sigma_1,\sigma_2,\sigma_3)\rightarrow(\pm\sigma_1,\pm\sigma_2,\pm\sigma_3)$ yield other polynomial roots of the related multi-valued system (\ref{cubesys}).
Note that although this transformation group contains eight elements, consistent with our earlier claim it only leads to four different values of $\thb{v}$ due to the symmetry $(\sigma_1,\sigma_2,\sigma_3)\rightarrow(-\sigma_1,-\sigma_2,-\sigma_3)$ of (\ref{cubesol}).

\section{Multi-soliton solutions}\label{SS}
The consistency on the cube provides immediately an auto-B\"acklund transformation for the equation defined by (\ref{model}):
\begin{equation}
\cQ_{p,k}(v,\wt{v},\wb{v},\wt{\wb{v}})=0, \quad \cQ_{q,k}(v,\wh{v},\wb{v},\wh{\wb{v}})=0.\label{bt}
\end{equation}
This system determines a new solution $\wb{v}=\wb{v}(n,m)$ from old solution $v=v(n,m)$, $k$ is the B\"acklund parameter.
The B\"acklund equations determining $\wb{v}$ are biquadratic correspondences associated with edges of the lattice.

That the equation defined by (\ref{model}) admits the constant solution $v=c$ can be seen by inspection, and it is natural to ask for the multi-soliton solution obtained by subsequent iteration of the B\"acklund transformation. 
At the first step of the B\"acklund chain the equations (\ref{bt}) are compatible autonomous symmetric biquadratic correspondences whose solution is standard.
Superposition can then be used to obtain the next few solutions.
Comparison with solutions obtained in \cite{hir3} motivates a slight generalisation of Hirota's original substitution and leads to the following $N$-soliton solution for model (\ref{model}):
\begin{equation}
v = c \frac{f f^{+-}}{f^+f^-}.\label{ss}
\end{equation}
Here $f$ and its counterparts are standard Hirota-type polynomials
\begin{equation}
\eqalign{
f:=\sum_{I\subseteq\{1\ldots N\}}Y_I\prod_{i\in I}\rho_i,\quad
f^{+-}:=\sum_{I\subseteq\{1\ldots N\}}Y_I\prod_{i\in I}a_i^+a_i^-\rho_i,\\
f^{+}:=\sum_{I\subseteq\{1\ldots N\}}Y_I\prod_{i\in I}a_i^{+}\rho_i,\quad
f^{-}:=\sum_{I\subseteq\{1\ldots N\}}Y_I\prod_{i\in I}a_i^{-}\rho_i,
}\label{fdef}
\end{equation}
where constants $Y_I, I\subseteq\{1\ldots N\}$ and $a_1^{\pm}\ldots a_N^{\pm}$ are defined as
\begin{equation}
Y_I:=\prod_{i,j\in I, i<j}\left(\frac{k_i^*-k_j^*}{k_i^*+k_j^*}\right)^2, \quad a_i^{\pm}:=\frac{1\pm c^2-k_i^*}{1\pm c^2+k_i^*},\label{Yadef}
\end{equation}
functions $\rho_1\ldots \rho_N$ are defined as
\begin{equation}
\rho_i:=\rho_{i,0}\left(\frac{p_*-k_i^*}{p_*+k_i^*}\right)^n\left(\frac{q_*-k_i^*}{q_*+k_i^*}\right)^m,\label{rhodef}
\end{equation}
and parameters $p_*$, $q_*$ and $k_1^*\ldots k_N^*$ are related to the lattice and B\"acklund parameters respectively by equations
\begin{equation}
\fl\quad p_*^2=1+pc^2+c^4, \quad q_*^2=1+qc^2+c^4, \quad (k_i^*)^2=1+k_ic^2+c^4, \quad i\in\{1\ldots N\}.\label{dud}
\end{equation}
Constants $\rho_{1,0}\ldots\rho_{N,0}$ and parameters $k_1^*\ldots k_N^*$ can be chosen freely in the solution.
Only the solution is given here, we omit recalling standard methods required to verify it (cf. \cite{ko,nah,hz}).
By inspecting (\ref{fdef}), (\ref{Yadef}), (\ref{rhodef}) and (\ref{dud}) above it can be seen that the specialisation $\{p,q\}=\{+2,-2\}$ returns solution (\ref{ss}) to the one obtained originally by Hirota:
\begin{equation}
v=c\frac{f\th{f}}{\wt{f}\wh{f}},
\end{equation}
which is consistent with observation (\ref{reduced}) made earlier.

\section{Relationship with Hirota's discrete KdV equation}\label{HM}
In this section model (\ref{model}) will be used to obtain a B\"acklund transformation for the equation due to Hirota \cite{hir3}.
The method relies on the multidimensional consistency of the associated single-valued vertex-bond system (\ref{dvdef}), (\ref{quadsys}), combined with property (\ref{reduced}).
Note that the transformation obtained by this method should be identified with the bilinear B\"acklund transformation given in \cite{hir3}.

The specialisation $p=-q=2$ reduces the system (\ref{dvdef}), (\ref{quadsys}) to
\begin{eqnarray}
\sigma_1 = (-1)^{n+m}(1+v\wt{v}), \quad \sigma_2 = (-1)^{n+m}(1-v\wh{v}),\label{redge}\\
v-\th{v} = \frac{1}{\wh{v}} - \frac{1}{\wt{v}},\label{hirota}
\end{eqnarray}
up to transformations $(\sigma_1,\sigma_2)\rightarrow(\pm\sigma_1,\pm\sigma_2)$, in particular recovering Hirota's equation (\ref{hirota}).
With the same specialisation the B\"acklund transformation for (\ref{dvdef}), (\ref{quadsys}), which is inherent from its three-dimensional consistency, leads to the following B\"acklund transformation for (\ref{hirota}):
\begin{eqnarray}
\eqalign{
v(v\wt{v}-\wb{v}\bt{v})(r-2)=2(v-\bt{v})(1+v\wt{v})(\sigma_3-1-v\wb{v}),\\
\wt{v}(v\wt{v}-\wb{v}\bt{v})(r-2)=2(\wb{v}-\wt{v})(1+v\wt{v})(\wt{\sigma}_3+1+\wt{v}\bt{v}),
}\label{hbt1}\\
\eqalign{
v(v\wh{v}-\wb{v}\wh{\wb{v}})(r+2)=2(v-\wh{\wb{v}})(1-v\wh{v})(\sigma_3-1+v\wb{v}),\\
\wh{v}(v\wh{v}-\wb{v}\wh{\wb{v}})(r+2)=2(\wb{v}-\wh{v})(1-v\wh{v})(\wh{\sigma}_3+1-\wh{v}\wh{\wb{v}}),
}\label{hbt2}
\end{eqnarray}
where substitution $\sigma_3\rightarrow (-1)^{n+m}\sigma_3$ has also been made in order to obtain the transformation in autonomous form.

For a fixed function $v=v(n,m)$ the systems (\ref{hbt1}) and (\ref{hbt2}) define bi-rational mappings $(\wb{v},\sigma_3)\mapsto(\bt{v},\wt{\sigma}_3)$ and $(\wb{v},\sigma_3)\mapsto(\wh{\wb{v}},\wh{\sigma}_3)$ which preserve the relation $\sigma_3^2=1+rv\wb{v}+v^2\wb{v}^2$, and which are associated with edges of the lattice in the $n$ and $m$ directions respectively.
The constraint (\ref{hirota}) on $v$ is sufficient for compatibility of the mappings, and they determine a new function $\wb{v}=\wb{v}(n,m)$ that again satisfies (\ref{hirota}).
The free parameter $r$ is the B\"acklund parameter.

This B\"acklund transformation determines not only $\wb{v}$ but also the auxiliary variable $\sigma_3$, which then plays a role in the superposition principle.
The superposition principle itself is obtained from the four-dimensional consistency of the system (\ref{dvdef}), (\ref{quadsys}):
let $(\wb{v},\sigma_3)$ and $(\wg{v},\sigma_4)$ be obtained by the B\"acklund transformation (\ref{hbt1}), (\ref{hbt2}) from a solution $v$ of (\ref{hirota}) with the corresponding B\"acklund parameters being $r$ and $s$.
Then the equation
\begin{equation}
\cF_{r,s}(v,\wb{v},\wg{v},\wg{\wb{v}},\sigma_3\sigma_4)=0\label{svsp}
\end{equation}
determines a new solution $\wg{\wb{v}}=\wg{\wb{v}}(n,m)$ of (\ref{hirota}) related to $v$ by the composition of the two transformations.
By using edge relations $\sigma_3^2=1+rv\wb{v}+v^2\wb{v}^2$ and $\sigma_4^2=1+sv\wg{v}+v^2\wg{v}^2$ the superposition principle (\ref{svsp}) can be expressed without the auxiliary variables as $\cQ_{r,s}(v,\wb{v},\wg{v},\wg{\wb{v}})=0$. 
Thus the full parameter-dependent model (\ref{model}) is recovered as superposition principle for the B\"acklund transformation (\ref{hbt1}), (\ref{hbt2}) of Hirota's equation (\ref{hirota}).

\section{Discussion: multi-quadratic models}\label{DIS}
To conclude we discuss further the discriminant factorisation property and method of introducing auxiliary variables which was used in Section \ref{DF}.

\subsection*{Weierstrass biquadratic}
The motivating example for the technique is classical theory of the Weierstrass biquadratic equation:
\begin{equation}
\eqalign{
\fl\quad (u\wt{u}+\wt{u}p+pu-e_1e_2-e_2e_3-e_3e_1)^2\\-4(u+\wt{u}+p-e_1-e_2-e_3)(u\wt{u}p-e_1e_2e_3) = 0.
}\label{wb}
\end{equation}
The single-valued system equivalent to the quadratic model (\ref{wb}) is usually written as
\begin{equation}
\fl\quad\frac{1}{4}\!\left(\frac{U-P}{u-p}\right)^2=u+\wt{u}+p-e_1-e_2-e_3,\quad
(\wt{u}-p)U+(u-\wt{u})P=(p-u)\wt{U},\label{wsv}
\end{equation}
where the auxiliary variable and parameter satisfy
\begin{equation}
U^2 = 4(u-e_1)(u-e_2)(u-e_3), \quad P^2 = 4(p-e_1)(p-e_2)(p-e_3).\label{waux}
\end{equation}
System (\ref{wsv}) defines a single-valued mapping $(u,U)\mapsto(\wt{u},\wt{U})$ which preserves the algebraic relation satisfied by the auxiliary variable in (\ref{waux}).
In this setting the analogue of multidimensional consistency is the well-known commutativity between the correspondence (\ref{wb}), or the associated single-valued mapping (\ref{wsv}), and its counterpart with different choice of parameter $p$.

\subsection*{Adler-Veselov model and the Kassotakis-Niesporski Idea-system variables}
The multi-quadratic model (\ref{avmodel}) was re-discovered in \cite{kn} through an associated single-valued {\it Idea system} involving variables only on lattice edges. 
This led in particular to a single-valued vertex-bond system equivalent to (\ref{avmodel}), and we wish to briefly show how the same system can be recovered by the constructive method based on discriminant factorisation.

The discriminant, with respect to variable $\th{u}$, of polynomial equation (\ref{avmodel}):
\begin{equation}
16(p-q)^2(\wt{u}-\wh{u})^2(u+\wt{u}+p)(u+\wh{u}+q),
\end{equation}
motivates introduction of the edge variables
\begin{equation}
\sigma_1^2=u+\wt{u}+p, \quad \sigma_2^2=u+\wh{u}+q,\label{avedge}
\end{equation}
which are exactly the variables of the associated {\it Idea system}.
As in Section \ref{DF} a single-valued vertex-bond system replaces (\ref{avmodel}) after introduction of the edge variables (\ref{avedge}), in this case it can be written as follows:
\begin{equation}
\eqalign{
(p+q)(u-\th{u}) - (p-q)(\wt{u}-\wh{u}) = (u-\th{u})(2\sigma_1\sigma_2-2u-\wt{u}-\wh{u}),\\
(p+q)(\wt{u}-\wh{u}) - (p-q)(u-\th{u}) = (\wt{u}-\wh{u})(2\sigma_1\wt{\sigma}_2-2\wt{u}-u-\th{u}),\\
(p+q)(\wh{u}-\wt{u}) - (p-q)(\th{u}-u) = (\wh{u}-\wt{u})(2\wh{\sigma}_1\sigma_2-2\wh{u}-\th{u}-u).
}\label{avsvm}
\end{equation}
Note that the {\it Idea system} itself, which may be written in the form
\begin{equation}
\sigma_1+\wh{\sigma}_1 = \sigma_2+\wt{\sigma}_2, \quad \sigma_1\wh{\sigma}_1-p=\sigma_2\wt{\sigma}_2-q,
\end{equation}
is also a consequence of system (\ref{avsvm}) and edge relations (\ref{avedge}).
Direct calculation using the vertex-bond system (\ref{avsvm}) was used in \cite{kn} to verify consistency on the cube of (\ref{avmodel}), we give here the same formula in a slightly different form more similar to (\ref{cubesol}):
\begin{equation}
\fl\quad\eqalign{
&(\wb{u}+\thb{u}-\wt{u}-\wh{u})r(p-q)^2-(\wb{u}-\thb{u})(2\sigma_1\sigma_2-2u-\wt{u}-\wh{u})(r-p)(q-r)\\
+&(\wt{u}+\thb{u}-\wh{u}-\wb{u})p(q-r)^2-(\wt{u}-\thb{u})(2\sigma_2\sigma_3-2u-\wh{u}-\wb{u})(p-q)(r-p)\\
+&(\wh{u}+\thb{u}-\wb{u}-\wt{u})q(r-p)^2-(\wh{u}-\thb{u})(2\sigma_3\sigma_1-2u-\wb{u}-\wt{u})(q-r)(p-q)=0,}\label{avsol}
\end{equation}
where $\sigma_3^2=u+\wb{u}+r$.
Note that the symmetry $(\sigma_1,\sigma_2,\sigma_3)\rightarrow(-\sigma_1,-\sigma_2,-\sigma_3)$ of (\ref{avsol}) means that initial data $\{u,\wt{u},\wh{u},\wb{u}\}$ determines only four distinct values of $\thb{u}$.

\subsection*{The discrete CKP equation}
The symmetric discrete Darboux system was given by Schief in \cite{sch} as a single-valued system involving variables on quad faces of the cubic lattice $\mathbb{Z}^3$  (cf. also \cite{ds,man}).
Introduction of an associated function $\tau$ on vertices as a potential for the face variables led to the following equation defined by a multi-quadratic polynomial:
\begin{equation}
(\tau\thb{\tau}-\th{\tau}\wb{\tau}+\hb{\tau}\wt{\tau}-\bt{\tau}\wh{\tau})^2-4(\tau\hb{\tau}-\wh{\tau}\wb{\tau})(\wt{\tau}\thb{\tau}-\th{\tau}\bt{\tau})=0,\label{CKP}
\end{equation}
which was shown to be superposition principle for the CKP hierarchy, and therefore it should be consistent in multidimensions.
This was confirmed in calculations by Tsarev and Wolf in \cite{tw}.
Here we observe that a single-valued system can be constructed from (\ref{CKP}) due to it having the discriminant factorisation property.
The discriminant, with respect to $\thb{\tau}$, of the multi-quadratic defining polynomial in (\ref{CKP}) is simply
\begin{equation}
16(\wt{\tau}\wh{\tau}-\tau\th{\tau})(\wh{\tau}\wb{\tau}-\tau\hb{\tau})(\wb{\tau}\wt{\tau}-\tau\bt{\tau}),
\end{equation}
and similar expressions hold for discriminants with respect to other variables due to (slightly hidden) symmetry of this polynomial.
The discriminants motivate introduction of new variables, this time on quad faces:
\begin{equation}
\sigma_{12}^2 = \wt{\tau}\wh{\tau}-\tau\th{\tau}, \quad 
\sigma_{23}^2 = \wh{\tau}\wb{\tau}-\tau\hb{\tau}, \quad 
\sigma_{13}^2 = \wb{\tau}\wt{\tau}-\tau\bt{\tau}.\label{face}
\end{equation}
The equivalent system with single-valued evolution that emerges is:
\begin{equation}
\eqalign{
\frac{\tau}{2}(\tau\thb{\tau}-\wt{\tau}\hb{\tau}-\wh{\tau}\bt{\tau}-\wb{\tau}\th{\tau})+\sigma_{12}\sigma_{23}\sigma_{13}+\wt{\tau}\wh{\tau}\wb{\tau}=0,\\
\frac{\wt{\tau}}{2}(\wt{\tau}\hb{\tau}-\tau\thb{\tau}-\th{\tau}\wb{\tau}-\bt{\tau}\wh{\tau})-\sigma_{12}\wt{\sigma}_{23}\sigma_{13}+\tau\th{\tau}\bt{\tau}=0,\\
\frac{\wh{\tau}}{2}(\wh{\tau}\bt{\tau}-\th{\tau}\wb{\tau}-\tau\thb{\tau}-\hb{\tau}\wt{\tau})-\sigma_{12}\sigma_{23}\wh{\sigma}_{13}+\th{\tau}\tau\hb{\tau}=0,\\
\frac{\wb{\tau}}{2}(\wb{\tau}\th{\tau}-\bt{\tau}\wh{\tau}-\hb{\tau}\wt{\tau}-\tau\thb{\tau})-\wb{\sigma}_{12}\sigma_{23}\sigma_{13}+\bt{\tau}\hb{\tau}\tau=0.
}
\label{ckpsvs}
\end{equation}
It contains mixed face and vertex variables unlike the original symmetric discrete Darboux formulation of this model which involves variables only on faces. 
In this respect this formulation is not so natural, however it demonstrates again the constructive method based on the discriminant factorisation property.
And for instance it allows direct verification of the multidimensional consistency of (\ref{CKP}) as follows.

First it is easily verified that (\ref{CKP}) is covariant, i.e. invariant under interchange of lattice directions, and this is inherited by (\ref{ckpsvs}).
It therefore remains to verify consistency of system (\ref{ckpsvs}) on (cubic) faces of a four-dimensional hypercube, it requires defining additional auxiliary variables
\begin{equation}
\sigma_{14}^2 = \wt{\tau}\wg{\tau}-\tau\wg{\wt{\tau}}, \quad 
\sigma_{24}^2 = \wh{\tau}\wg{\tau}-\tau\wg{\wh{\tau}}, \quad 
\sigma_{34}^2 = \wb{\tau}\wg{\tau}-\tau\wg{\wb{\tau}},\label{face2}
\end{equation}
and a new shift (grave) in a fourth lattice direction.
Direct calculation obtains the equation
\begin{equation}
\eqalign{
\fl\quad\frac{\tau^2}{2}(\tau\wg{\thb{\tau}}-\th{\tau}\wg{\wb{\tau}}-\hb{\tau}\wg{\wt{\tau}}-\bt{\tau}\wg{\wh{\tau}})+\sigma_{12}\sigma_{23}\sigma_{13}\wg{\tau} + \sigma_{23}\sigma_{34}\sigma_{24}\wt{\tau} + \sigma_{13}\sigma_{14}\sigma_{34}\wh{\tau} +\sigma_{12}\sigma_{24}\sigma_{14}\wb{\tau}\\ + \sigma_{12}\sigma_{34}\sigma_{13}\sigma_{24}+\sigma_{13}\sigma_{24}\sigma_{14}\sigma_{23}+\sigma_{14}\sigma_{23}\sigma_{12}\sigma_{34} + \wt{\tau}\wh{\tau}\wb{\tau}\wg{\tau}=0
}\label{4dc}
\end{equation}
which determines $\wg{\thb{\tau}}$ symmetrically from initial data on the hypercube.
Note that consistent with the observation made in \cite{tw}, initial data $\{\tau,\wt{\tau},\wh{\tau},\wb{\tau},\wg{\tau},\th{\tau},\bt{\tau},\wg{\wt{\tau}},\hb{\tau},\wg{\wh{\tau}},\wg{\wb{\tau}}\}$ determines only eight values of $\wg{\thb{\tau}}$.
This can be seen here because although the transformation group $(\sigma_{12},\sigma_{13},\sigma_{14},\sigma_{23},\sigma_{24},\sigma_{34})\rightarrow(\pm\sigma_{12},\pm\sigma_{13},\pm\sigma_{14},\pm\sigma_{23},\pm\sigma_{24},\pm\sigma_{34})$ contains 64 elements, the equation (\ref{4dc}) has four symmetries of the form $(\sigma_{12},\sigma_{13},\sigma_{14},\sigma_{23},\sigma_{24},\sigma_{34})\rightarrow(-\sigma_{12},-\sigma_{13},-\sigma_{14},\sigma_{23},\sigma_{24},\sigma_{34})$ which generate a sub-group of eight elements. 

\subsection*{Conclusion}
It has been observed here how a discriminant factorisation property can assuage the apparent awkwardness associated with the multi-valued evolution of multi-quadratic discrete models.
Moreover, its ubiquity amongst the known integrable examples suggests to investigate the extent to which this simple feature is connected to their integrability.

\ack
I would like to express my sincerest thanks to Maciej Nieszporski for many interesting and helpful discussions and for pointing out several of the references.
This research was funded by Australian Research Council Discovery Grant DP 110104151.

\end{document}